\documentclass[apjl]{emulateapj}
\usepackage{amsmath}
\usepackage{lineno}

\begin{document}
\bibliographystyle{apj}

\title{Do close-in giant planets orbiting evolved stars prefer eccentric orbits?}

\author{Samuel K.\ Grunblatt\altaffilmark{1,*}}

\shorttitle{Eccentric Hot Jupiters Around Red Giants}
\shortauthors{Grunblatt et al.}

\author{Daniel Huber\altaffilmark{1,2,3,4}}
\author{Eric Gaidos\altaffilmark{5}}
\author{Eric D.\ Lopez\altaffilmark{6}}
\author{Thomas Barclay\altaffilmark{6}}
\author{Ashley Chontos\altaffilmark{1}}
\author{Evan Sinukoff\altaffilmark{1,7}}
\author{Vincent Van Eylen\altaffilmark{8}}
\author{Andrew W.\ Howard\altaffilmark{7}}
\author{Howard T. Isaacson\altaffilmark{9}}
%

\altaffiltext{1}{Institute for Astronomy, University of Hawaii,
2680 Woodlawn Drive, Honolulu, HI 96822, USA}
\altaffiltext{2}{Sydney Institute for Astronomy (SIfA), School of Physics, University of 
Sydney, NSW 2006, Australia}
\altaffiltext{3}{SETI Institute, 189 Bernardo Avenue, Mountain View, CA 94043, USA}
\altaffiltext{4}{Stellar Astrophysics Centre, Department of Physics and Astronomy, 
Aarhus University, Ny Munkegade 120, DK-8000 Aarhus C, Denmark}
\altaffiltext{5}{Department of Geology $\&$ Geophysics, University of
Hawaii at Manoa, Honolulu, Hawaii 96822, USA} 
\altaffiltext{6}{NASA Goddard Space Flight Center, Greenbelt, MD 20771, USA}
\altaffiltext{7}{California Institute of Technology, Pasadena, CA 91125, USA}
\altaffiltext{8}{Leiden Observatory, Leiden University, posts 9513, 2300RA Leiden, The Netherlands}
\altaffiltext{9}{Department of Astronomy, UC Berkeley, Berkeley, CA 94720, USA}

\altaffiltext{*}{skg@ifa.hawaii.edu}

\begin{abstract}


The NASA \emph{Kepler} and \emph{K2} Missions have recently revealed a population of transiting giant planets orbiting moderately evolved, low-luminosity red giant branch stars. Here, we present radial velocity measurements of three of these systems, revealing significantly non-zero orbital eccentricities in each case. Comparing these systems with the known planet population suggests that close-in giant planets around evolved stars tend to have more eccentric orbits than those around main-sequence stars. We interpret this as tentative evidence that the orbits of these planets pass through a transient, moderately eccentric phase where they shrink faster than they circularize due to tides raised on evolved host stars. Additional radial velocity measurements of currently known systems, along with new systems discovered by the recently launched NASA \emph{TESS} mission, may constrain the timescale and mass dependence of this process.


\end{abstract}

\section{Introduction}


The NASA \emph{Kepler} mission has discovered thousands of extrasolar planets, allowing populations of planets orbiting different types of stars to be compared \citep{howard2012, petigura2013, dressing2015, santerne2016, fulton2017, vansluijs2018}. However, the population of planets around evolved stars remained poorly described because so few have been discovered to date, particularly at orbital distances of 0.5 AU or less \citep{sato2005, johnson2010, lillo-box2014, barclay2015,jones2016}. 

It has been suggested that the planet population of evolved stars should look quite different from their main sequence counterparts due to dynamical interactions driven by stellar evolution \citep{veras2016}. Accelerated angular momentum exchange should cause the planet to spiral in to the host star \citep{zahn1977, hut1981, macleod2018}. This results in a scenario where orbital decay happens faster than circularization, producing a population of transient, moderately eccentric close-in planets around evolved stars that are not seen around main sequence stars \citep{villaver2009, villaver2014}. The increase in planetary heating from both elevated stellar irradiation and tides raised on the planet will likely also cause inflation of these planets at late times \citep{bodenheimer2001, lopez2016}.

Two well-characterized, close-in inflated giant planets orbiting moderately evolved, or low-luminosity red giant branch stars, K2-97b and K2-132b, were recently discovered by the \emph{K2} extension to the \emph{Kepler} mission \citep{grunblatt2016, grunblatt2017}. Here, we report new radial velocity (RV) measurements of these planets, as well as RV measurements of a previously validated planet orbiting an evolved star observed by the original \emph{Kepler} mission, Kepler-643 \citep{huber2013, morton2016}. These measurements allow us to constrain the orbital eccentricities of these planets, which motivate an investigation of the orbital eccentricities of the population of planets around giant stars compared to dwarf stars.



\begin{deluxetable*}{llllllll}
\tabletypesize{\scriptsize}
\tablecaption{Close-In Giant Planets Orbiting Giant Stars \label{tbl-star}}
\tablewidth{\linewidth}
\tablehead{
\colhead{Name} & \colhead{Mass} & \colhead{Radius} & \colhead{Semimajor Axis} & \colhead{Eccentricity} & \colhead{Stellar Mass} & \colhead{Stellar Radius} & \colhead{Reference}
}

\startdata
K2-132b & 0.49 $\pm$ 0.06 M$_\mathrm{J}$ & 1.30 $\pm$0.07 R$_\mathrm{J}$ & 0.086 AU & 0.36 $\pm$ 0.06 & 1.08 $\pm$ 0.08 M$_\odot$ &  3.85 $\pm$ 0.13 R$_\odot$ & 1, this work\\
K2-97b & 0.48 $\pm$ 0.07 M$_\mathrm{J}$ & 1.31 $\pm$0.11 R$_\mathrm{J}$ & 0.081 AU & 0.22 $\pm$ 0.08 & 1.16 $\pm$ 0.12 M$_\odot$ & 4.20 $\pm$ 0.14 R$_\odot$ & 1, this work\\
K2-39b & 0.125 $\pm$ 0.014 M$_\mathrm{J}$ & 0.51 $\pm$ 0.06 R$_\mathrm{J}$ & 0.057 AU & 0.15 $\pm$ 0.08 & 1.19 $\pm$ 0.08 M$_\odot$ &  2.93 $\pm$ 0.21 R$_\odot$ & 2\\
Kepler-643b & 1.01 $\pm$ 0.20 M$_\mathrm{J}$ & 1.14 $\pm$ 0.05 R$_\mathrm{J}$ & 0.126 AU & 0.37 $\pm$ 0.06 & 1.15 $\pm$ 0.12 M$_\odot$ & 2.69 $\pm$ 0.11 R$_\odot$ & 3, 4, this work\\
Kepler-91b & 0.81 $\pm$ 0.18 M$_\mathrm{J}$ & 1.37 $\pm$ 0.07 R$_\mathrm{J}$ & 0.0731 AU & 0.04$^{+0.06}_{-0.02}$ & 1.31 $\pm$ 0.1 M$_\odot$ & 6.30 $\pm$ 0.16 R$_\odot$ & 5\\
HD 102956b & 0.96 $\pm$ 0.05 M$_\mathrm{J}$ & non-transiting & 0.081 AU & 0.05 $\pm$ 0.03 & 1.70 $\pm$ 0.11 M$_\odot$ &  4.4 $\pm$ 0.1 R$_\odot$ & 6\\
TYC3667-1280-1b & 5.4 $\pm$ 0.4 M$_\mathrm{J}$ & non-transiting & 0.21 AU & 0.04$^{+0.04}_{-0.02}$ & 1.87 $\pm$ 0.17 M$_\odot$ &  6.26 $\pm$ 0.86 R$_\odot$ & 7

\enddata
\tablecomments{Reference key: 1. \citet{grunblatt2017}, 2. \citet{petigura2017b}, 3. \citet{huber2013}, 4. \citet{morton2016}, 5. \citet{barclay2015}, 6. \citet{johnson2010}, 7. \citet{niedzielski2016}. }
\end{deluxetable*}


\section{Observations}

RV measurements of K2-97, K2-132, and Kepler-643 were obtained between 2016 January 27 and 2018 February 1 using the High Resolution Echelle Spectrometer (HIRES) on the Keck-I Telescope at the Maunakea Observatory in Hawaii. Individual measurements and orbit solutions are shown in Figure 1. All RV spectra were obtained through an iodine gas cell. In order to constrain orbital parameters, we fit the radial velocity data using the publicly available software package \texttt{RadVel} \citep{fulton2018}. The orbital period of the planets were fixed to published values from transit measurements \citep{morton2016, grunblatt2017}, while we fit for the semi-amplitude, phase,and modified eccentricity parameters of the orbit \citep{eastman2013}. We also fit for an RV jitter term for our measurements and obtained a value between 5-10 m s$^{-1}$ for all systems studied here. We adopted the same method for determining RVs as described in \citet{butler1996}. 


\begin{figure}[ht!]
\epsscale{1.2}
\plotone{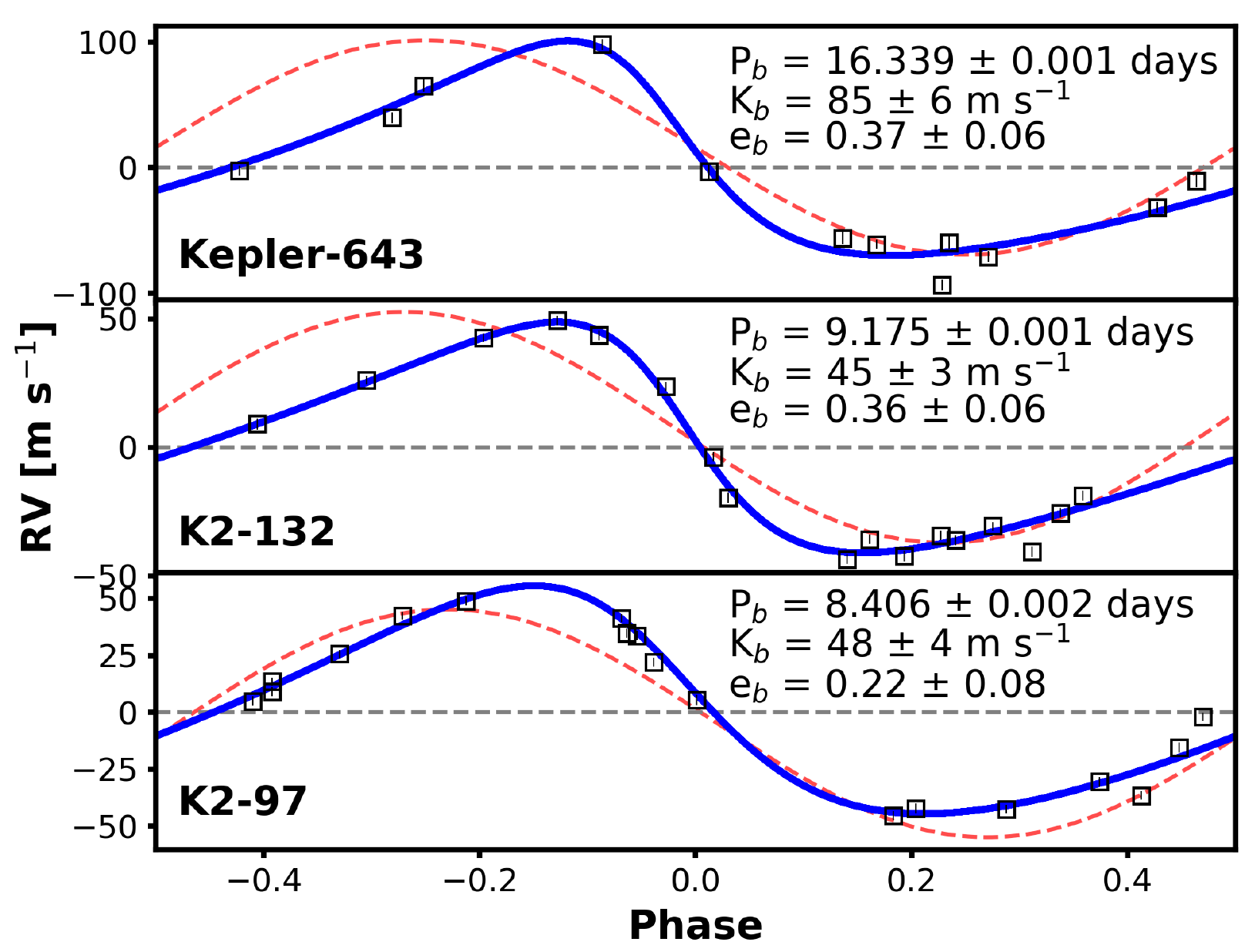}
\caption{ Keck/HIRES radial velocity observations of Kepler-643 (top), K2-132 (center) and K2-97 (bottom), three systems where close-in giant planets orbit evolved stars. All orbits display moderate eccentricities between 0.2 and 0.4. The planets appear to follow a trend, where those on longer orbits are more eccentric than those orbiting their host star more closely. Circular orbits are shown as red dotted lines for reference. \label{rvs}}
\end{figure}

Since RV measurements are not usually taken at regular time intervals, data sampling is often uneven and thus introduces orbital parameter biases, potentially inflating eccentricities beyond their true value \citep{eastman2013}. To ensure that our measured eccentricities are robust, we produced 100 artificial RV datasets of circular orbits for each system, with equivalent orbital periods, semi-amplitudes, and random scatter as measured in our real data, taken at the same times as our real measurements. We then recovered an orbit from each artificial dataset using the same techniques given for our real RV data.  We find that the distribution of eccentricities recovered from fitting the artificial datasets is consistent with zero in all cases. For all best fit orbit solutions for the simulated, $e$=0 orbit generated data, we do not recover an eccentricity of greater than 0.1. We therefore conclude that the eccentricities found by our analysis are not due to sparse sampling of our RV measurements.



\section{Eccentricity Distributions Around Evolved Stars}

\begin{figure*}[ht!]
\epsscale{1.15}
\plottwo{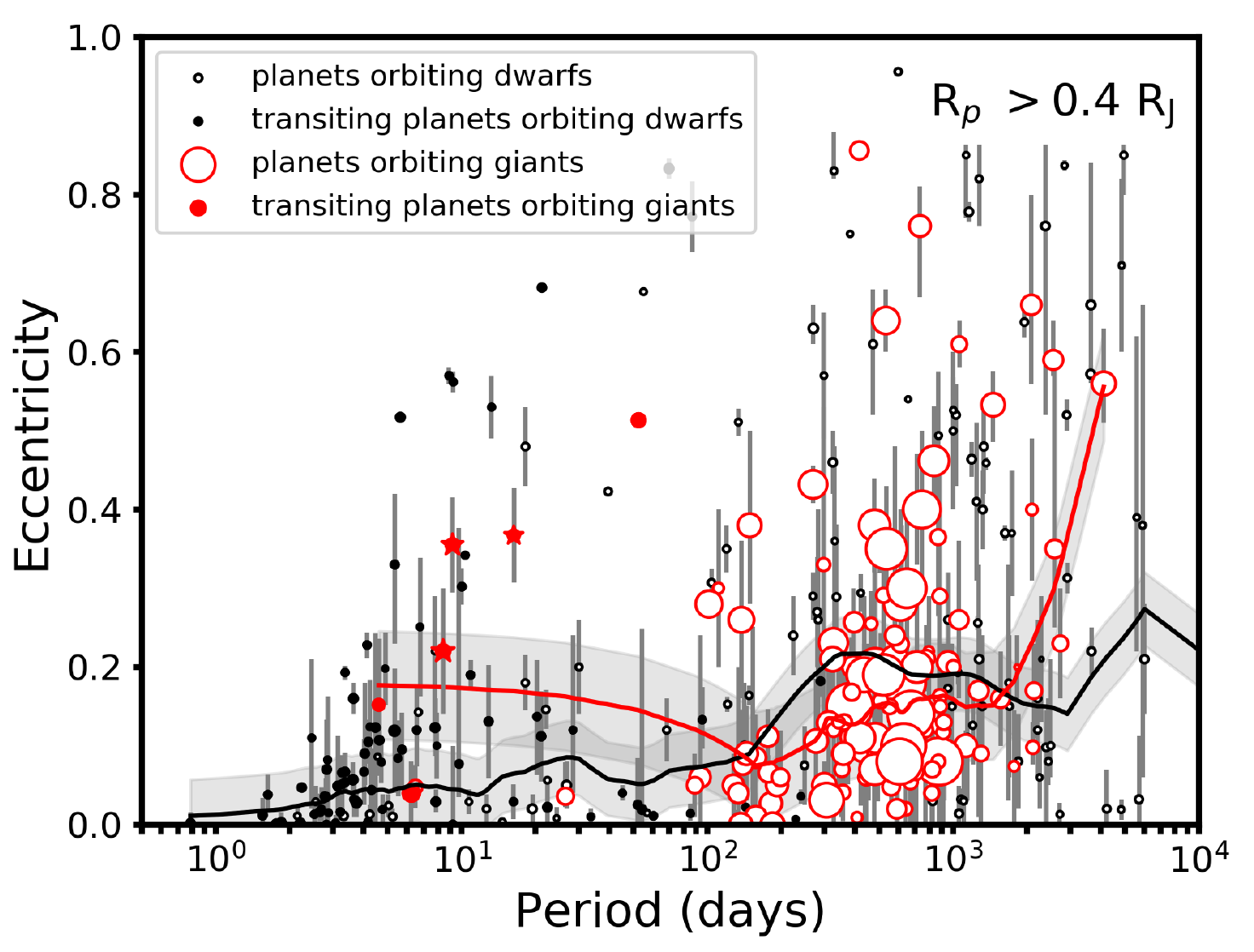}{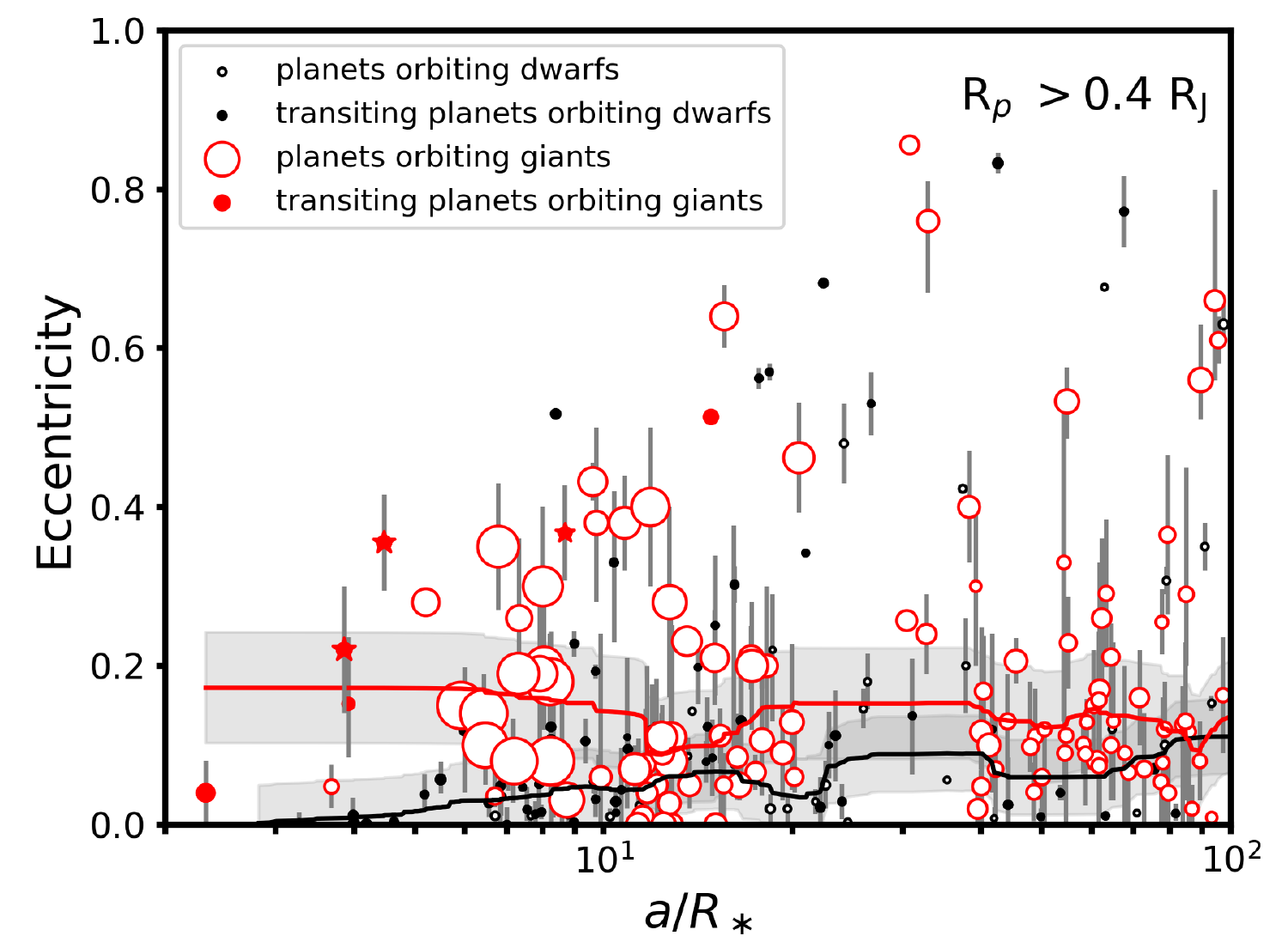}
\caption{Left: Orbital period versus eccentricity for all giant ($>$0.4 R$_\mathrm{J}$) planets with published eccentricities orbiting giant and dwarf stars. Stellar radius scales with the size of the points; planets orbiting giant stars are shown in red, while planets orbiting dwarfs are shown in black. The systems with eccentricities measured in this study are highlighted as red stars. A locally weighted regression of the eccentricities of are shown by the solid black and red lines for the dwarf and giant star populations, respectively. Right: Same as left, except with $a$/R$_*$ on the x-axis. \label{fig2}}
\vspace{0.84cm}
\end{figure*}

\begin{figure}[ht!]
\epsscale{1.15}
\plotone{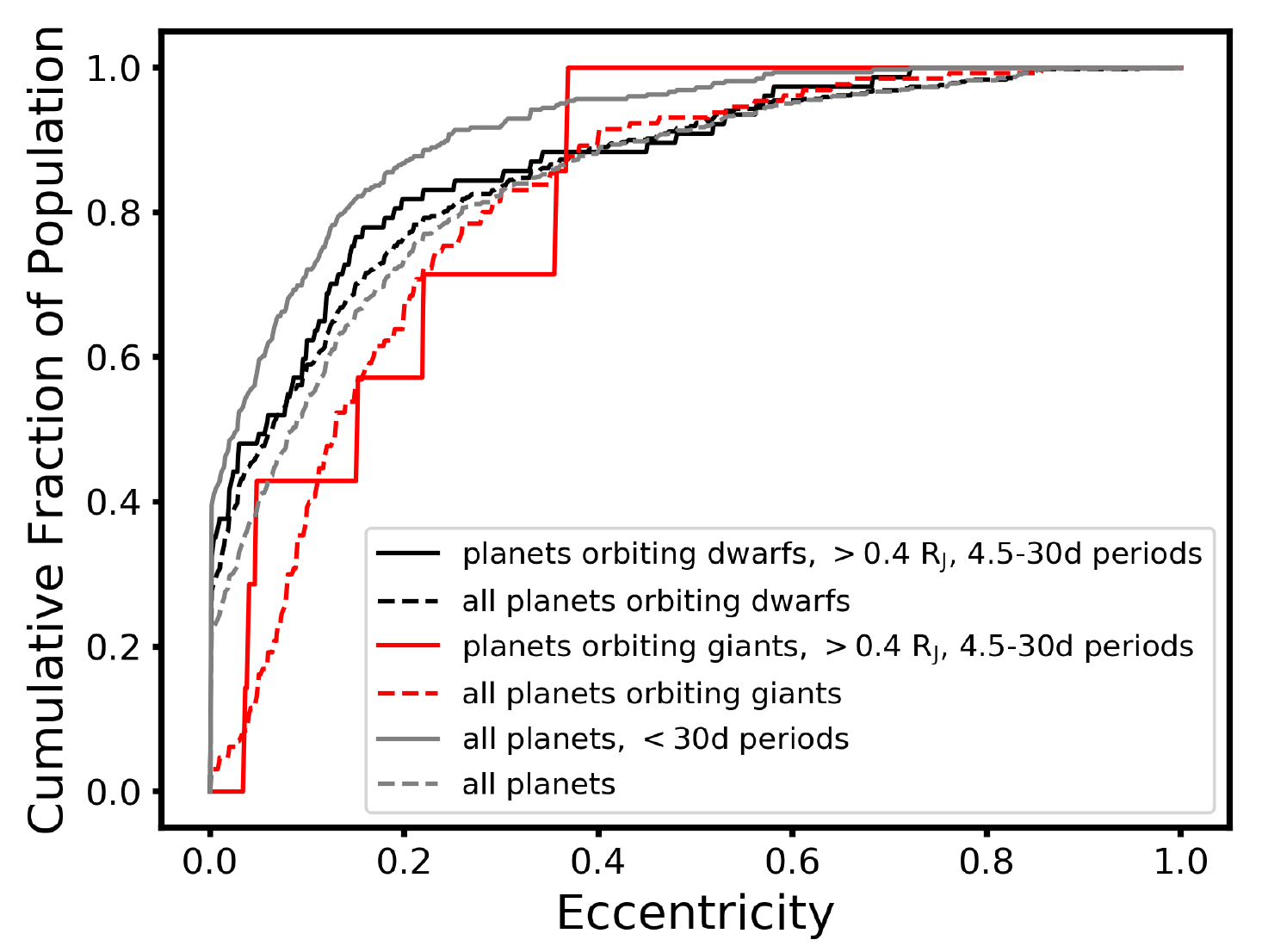}
\caption{Cumulative eccentricity distributions of different populations of planets. Planets orbiting giant stars (red lines), particularly at periods of 30 days or less, display a preference for moderate eccentricities not seen in dwarf star systems (black lines). \label{fig3}}
\vspace{0.84cm}
\end{figure}


Figure 2 illustrates the population of known giant planets with published eccentricities orbiting giant stars as well as the equivalent planet population orbiting dwarfs in the orbital period and eccentricity plane (left) and the $a$/R$_*$ and eccentricity plane (right). Planets are designated as giants if $R_p$ $>$ 0.4 $R_\mathrm{J}$. 419 dwarf star systems and 136 giant star systems with constrained eccentricities listed in the NASA Exoplanet Archive are included in our figure \citep{akeson2013}. Transiting systems are shown as filled circles, while non-transiting systems are shown as empty circles. For non-transiting systems, planet radii were estimated using the mass-radius relations of \citet{chen2017}. Distinctions as giant or dwarf star systems were made using the physically motivated boundaries in effective temperature and surface gravity described in \citet{huber2016}. Stellar parameters have been taken from the NASA Exoplanet Archive, and individual sources for all known close-in giant planets with published eccentricities orbiting giant stars are listed in Table 1. Our new RV measurements give tentative evidence that the dwarf and giant system eccentricity distributions are inconsistent at periods $\lesssim$ 50 days and $a$/R$_*$ $\lesssim$ 10.

Figure 3 illustrates the cumulative distributions of eccentricities for various different planetary system samples analyzed here. When considering planets of all sizes, close-in planets show a tendency for low eccentricities. However, this preference is not as strong when considering only giant planets, likely due to trends related to planet multiplicity \citep{vaneylen2015, xie2016}. Remarkably, comparing the population of giant planets orbiting at $\lesssim$ 50 day orbital periods as well as all known planets around giant stars (red lines) to the equivalent planet population orbiting dwarf stars (black lines) illustrates a stronger preference for moderate eccentricities in giant star systems than is seen in dwarf star systems.   


To evaluate the significance of the difference between the dwarf and giant star planet populations, we compared the median eccentricities for both populations (see Figures 2 and 3). We restrict our analysis to giant ($>$ 0.4 $R_\mathrm{J}$) planets with orbital periods between 4.5 and 30 days and published eccentricity constraints. This ensures that all planets compared here could have been detected around both dwarf and low-luminosity red giant branch stars observed by \emph{K2}. Furthermore, this sample includes the closest-in known transiting planets orbiting evolved stars while rejecting the shortest period dwarf system planets, which likely would be engulfed by evolved stars due to their large sizes. It also minimizes biases due to planets found in surveys which were particularly well-suited to discovering short-period giant planets on circular orbits around dwarf stars \citep[{\it e.g.}, WASP,][]{pollacco2006}. Planets with published upper limits on eccentricity are treated as having circular orbits with error distributions that reach the listed upper limit at a 1-$\sigma$ confidence interval. We find a median eccentricity of 0.152$^{+0.077}_{-0.042}$ for close-in giant planets orbiting evolved stars, and a median eccentricity of 0.056$^{+0.022}_{-0.006}$ for close-in giant planets orbiting dwarfs. 


We also tested the sensitivity of these values to increasing the planet radius cut to $>$ 0.8 $R_\mathrm{J}$, as well as adjusting the inner period bound between 3-8 days, and the outer period bound between 25-80 days. We find that our statistics are only significantly affected by changing the inner period bound, driven by the small number of close-in planets known orbiting evolved stars. Thus, we choose bounds to include all known close-in planets orbiting evolved stars while minimizing the number of close-in planets around dwarf stars without an evolved counterpart population.



To further quantify the significance of the eccentricity dichotomy between the populations of giant planets orbiting dwarf and giant stars, we calculate the Anderson-Darling statistic, which is more robust to different-sized and small number distributions than similar tests such as the Kolmogorov-Smirnov statistic \citep{simpson1951, stephens1974}. We find that both samples are drawn from the same parent population in 6.3\% or fewer of cases. Adjusting our planet radius and period cuts, we find that both samples are drawn from the same parent population in 3.8\%--15.4\% or fewer of cases for all tested samples. This range is dominated by stochastic variation due to the small sample of evolved systems. 



As an additional test, we performed a Monte Carlo simulation in which we drew an equal number of eccentricity values from the eccentricity distributions of our bias-resistant sample of close-in giant planets orbiting dwarf stars and giant stars in 4.5-30 days. We find that after repeating this process one million times, the random sample of planets drawn from the dwarf star sample has a similar or higher median eccentricity than the planets orbiting giant stars in 5.7\% of cases, with a range of 4.1\% to 16.7\% for all period and radius ranges tested. We also performed the same test for the population of all close-in planets known around dwarf and giant stars, as well as all planets known around dwarf and giant stars, and find that the dwarf star sample has a similar or higher median eccentricity in 0.34\% and 10.6\% of cases, respectively. 

Thus, based on our statistical tests, we conclude that close-in, evolved star system planets display different eccentricity characteristics than close-in dwarf star system planets at a 1- to 2-$\sigma$ level. We note that this is a conservative estimate, as many early literature estimates of eccentricities for both types of systems may be biased toward higher eccentricities due to mischaracterization of systematic and astrophysical uncertainties \citep{eastman2013}. More recent RV studies, using newer analysis packages such as \texttt{RadVel}, account for this artificial bias. Reanalysis of RV measurements used to constrain the population of planetary eccentricities could remove this bias, but is beyond the scope of this Letter.

\section{Discussion}


The formation of close-in giant planets is commonly explained by three different hypotheses: {\it in situ} formation, disk migration, and tidal migration (see \citet{dawson2018} for a recent review). Populations of eccentric giant planets are generally viewed as evidence for tidal migration, as they cannot be explained by the other two prevailing mechanisms. Although these planets support tidal migration theory for close-in giant planet formation, we assert that unlike those around dwarf stars, these close-in giant planets are actively undergoing tidal migration, sped up by the late stage evolution of their host stars. An observed correlation between stellar host evolutionary state and long-period, planetary companions to close-in giant planet systems supports this \citep{lillo-box2016}. 



Models of the dynamical evolution of close-in giant planets can be strongly affected by the evolution of the host star \citep{villaver2009, villaver2014}. The timescale of this dynamical evolution is defined by the tidal interactions between the planet and its host star. Following the reasoning of \citet{villaver2014}, the eccentricity evolution of a planetary orbit will be dominated by planetary tides driving orbit circularization on the main sequence, and stellar tides driving tidal inspiral on the red giant branch. For example, assuming $Q_p = Q_* \sim$ 10$^6$, and using the equilibrium tide formulations of \citet{patra2017} derived from \citet{goldreich1966}, the timescale for orbit circularization for K2-97b is $\sim$5 Gyr, while the tidal inspiral timescale is $\lesssim$2 Gyr. This suggests orbital decay is driven more rapidly than eccentricity evolution as the stellar radius increases, producing a population of transient planets displaying moderate eccentricities at close-in orbits around evolved stars. Though these tidal timescale formulae do not account for planetary or stellar rotation or dynamical tides, these results are consistent with our observations.









\citet{villaver2014} also predict that more massive systems evolve more quickly toward lower eccentricities and semimajor axes. This is also tentatively supported by observations, as the most massive hosts in our sample also have the lowest eccentricity orbits (see Table 1). However, a larger sample of systems is needed to confirm this. Correlations between planet and star mass and composition and planetary orbital evolution have not yet been fully explored.    



Tidal interaction and migration has long been thought to cause radius inflation in gas-giant planets \citep{bodenheimer2001, storch2014}. Increased irradiation due to stellar evolution is also thought to be a source of planetary heating \citep{lopez2016}. Two of the close-in evolved planets with new RV measurements presented here, K2-97b and K2-132b, show signs of being significantly inflated relative to similar planets seen orbiting main sequence stars \citep{grunblatt2017}. 

To evaluate the dominant radius inflation mechanism for these planets, we follow the prescription for tidal heating given by \citet{miller2009} and \citet{dobbs-dixon2004}, and assume synchronous rotation of the planet and tidal quality factors $Q_p$ = 10$^4$ and $Q_*$  = 10$^6$, within an order of magnitude of observed and model constraints \citep{patra2017, Gallet2017}. We find that if the planets are actively circularizing, tidal evolution driven by the star can dominate planetary heating by an order of magnitude over irradiative mechanisms. Furthermore, tidal resonance locking may also greatly enhance tidal heating rates \citep{fuller2017}. Thus, planet radius inflation for these systems may be driven solely by tidal processes. 


However, a $Q_p$ value of 10$^4$ and $Q_* = 10^6$ would suggest the orbit circularization timescale is significantly shorter than the orbital decay timescale. In contrast, the observed eccentricities of these planet orbits suggests that orbit circularization and orbital decay are happening on similar timescales, implying $Q_* \sim Q_p$. This disagrees with predictions of $Q_*$ for evolved stars \citep{Gallet2017}. Furthermore, rotation and/or dynamical tides can drastically change these timescales and may even increase orbital eccentricity over time \citep{hut1981,fuller2017}. Determining the orbital evolution of evolved systems and causes of late stage planet inflation will require more in-depth characterization of the combined effect of increased irradiation and tidal energy dissipation on a larger sample of planets.

\section{Summary and Outlook}


The NASA \emph{Kepler} and \emph{K2} Missions have recently revealed a population of giant planets at small orbital separations around evolved stars. Here, we report radial velocity observations which show that a majority of these planets display moderate eccentricities, indicating a different evolutionary state for planets around giant stars than those orbiting main sequence stars. This late stage evolution is likely driven by the increase in size of the stellar radius and convective envelope, strongly increasing the angular momentum exchange between the star and the planet, causing the planet to circularize its orbit and spiral into the host star. These two components of orbital evolution must happen on timescales similar enough such that these migrating giant planets with moderate eccentricities appear to be relatively common around evolved stars \citep{villaver2014}. These planets will thus allow constraints on the determination of the tidal quality factors $Q_p$ and $Q_*$. Continued follow-up of low-luminosity red giant branch stars will allow estimation of close-in planetary occurrence around evolved stars (Grunblatt et al. 2018, {\it in prep}.), which will further constrain our understanding of planetary evolution and dynamical interactions within planetary systems.

Additional eccentricity constraints and more systems are needed in order to confirm the tentative result presented here. The NASA \emph{TESS} Mission, launched earlier this year, will observe two orders of magnitude as many evolved stars as \emph{Kepler} and \emph{K2}, likely resulting in over 100 planet detections around evolved stars \citep{sullivan2015, campante2016, barclay2018}. This detection of additional planets orbiting evolved stars will outline the diversity of all such systems, and the likelihood and timescale of planetary system disruption via stellar tides. With this information, we can investigate how quickly planets undergo orbital evolution around low-luminosity red giant branch stars, and at what point planets can no longer survive around giant stars, significantly distinguishing these systems from planet populations of main sequence stars.


\acknowledgements{The authors would like to thank John Livingston, Jim Fuller, Masanobu Kunitomo, Alessandro Brani, Amaury Triaud, Benjamin Pope, Lauren Weiss, Teruyuki Hirano, Travis Berger, Jessica Stasik, Connor Auge, Aaron Do, and Nader Haghighipour for helpful discussions. This research was supported by the NASA K2 Guest Observer Awards NNX16AH45G and NNX17AF76G to D.H.. D.H. acknowledges support by the National Aeronautics and Space Administration under Grant NNX14AB92G issued through the Kepler Participating Scientist Program. This research has made use of the Exoplanet Orbit Database and the Exoplanet Data Explorer at Exoplanets.org. This work was based on observations at the W. M. Keck Observatory granted by the University of Hawaii, the University of California, and the California Institute of Technology. We thank the observers who contributed to the measurements reported here and acknowledge the efforts of the Keck Observatory staff. We extend special thanks to those of Hawaiian ancestry on whose sacred mountain of Maunakea we are privileged to be guests.}


\begin{thebibliography}{}
\expandafter\ifx\csname natexlab\endcsname\relax\def\natexlab#1{#1}\fi

\bibitem[{{Barclay} {et~al.}(2015){Barclay}, {Endl}, {Huber}, {Foreman-Mackey},
  {Cochran}, {MacQueen}, {Rowe}, \& {Quintana}}]{barclay2015}
{Barclay}, T., {Endl}, M., {Huber}, D., {et~al.} 2015, \apj, 800, 46

\bibitem[{{Bodenheimer} {et~al.}(2001){Bodenheimer}, {Lin}, \&
  {Mardling}}]{bodenheimer2001}
{Bodenheimer}, P., {Lin}, D.~N.~C., \& {Mardling}, R.~A. 2001, \apj, 548, 466

\bibitem[{{Bowler} {et~al.}(2010){Bowler}, {Johnson}, {Marcy}, {Henry}, {Peek},
  {Fischer}, {Clubb}, {Liu}, {Reffert}, {Schwab}, \& {Lowe}}]{bowler2010}
{Bowler}, B.~P., {Johnson}, J.~A., {Marcy}, G.~W., {et~al.} 2010, \apj, 709,
  396

\bibitem[{{Butler} {et~al.}(1996){Butler}, {Marcy}, {Williams}, {McCarthy},
  {Dosanjh}, \& {Vogt}}]{butler1996}
{Butler}, R.~P., {Marcy}, G.~W., {Williams}, E., {et~al.} 1996, \pasp, 108, 500

\bibitem[{{Farihi} {et~al.}(2013){Farihi}, {G{\"a}nsicke}, \&
  {Koester}}]{farihi2013}
{Farihi}, J., {G{\"a}nsicke}, B.~T., \& {Koester}, D. 2013, Science, 342, 218

\bibitem[{{Fulton} {et~al.}(2018){Fulton}, {Petigura}, {Blunt}, \&
  {Sinukoff}}]{fulton2018}
{Fulton}, B.~J., {Petigura}, E.~A., {Blunt}, S., \& {Sinukoff}, E. 2018, ArXiv
  e-prints, arXiv:1801.01947

\bibitem[{{Fulton} {et~al.}(2017){Fulton}, {Petigura}, {Howard}, {Isaacson},
  {Marcy}, {Cargile}, {Hebb}, {Weiss}, {Johnson}, {Morton}, {Sinukoff},
  {Crossfield}, \& {Hirsch}}]{fulton2017}
{Fulton}, B.~J., {Petigura}, E.~A., {Howard}, A.~W., {et~al.} 2017, ArXiv
  e-prints, arXiv:1703.10375

\bibitem[{{Gillon} {et~al.}(2017){Gillon}, {Triaud}, {Demory}, {Jehin}, {Agol},
  {Deck}, {Lederer}, {de Wit}, {Burdanov}, {Ingalls}, {Bolmont}, {Leconte},
  {Raymond}, {Selsis}, {Turbet}, {Barkaoui}, {Burgasser}, {Burleigh}, {Carey},
  {Chaushev}, {Copperwheat}, {Delrez}, {Fernandes}, {Holdsworth}, {Kotze}, {Van
  Grootel}, {Almleaky}, {Benkhaldoun}, {Magain}, \& {Queloz}}]{gillon2017}
{Gillon}, M., {Triaud}, A.~H.~M.~J., {Demory}, B.-O., {et~al.} 2017, \nat, 542,
  456

\bibitem[{{Grunblatt} {et~al.}(2016){Grunblatt}, {Huber}, {Gaidos}, {Lopez},
  {Fulton}, {Vanderburg}, {Barclay}, {Fortney}, {Howard}, {Isaacson}, {Mann},
  {Petigura}, {Silva Aguirre}, \& {Sinukoff}}]{grunblatt2016}
{Grunblatt}, S.~K., {Huber}, D., {Gaidos}, E.~J., {et~al.} 2016, \aj, 152, 185

\bibitem[{{Grunblatt} {et~al.}(2017){Grunblatt}, {Huber}, {Gaidos}, {Lopez},
  {Howard}, {Isaacson}, {Sinukoff}, {Vanderburg}, {Nofi}, {Yu}, {North},
  {Chaplin}, {Foreman-Mackey}, {Petigura}, {Ansdell}, {Weiss}, {Fulton}, \&
  {Lin}}]{grunblatt2017}
{Grunblatt}, S.~K., {Huber}, D., {Gaidos}, E., {et~al.} 2017, \aj, 154, 254

\bibitem[{{Howard} {et~al.}(2012){Howard}, {Marcy}, {Bryson}, {Jenkins},
  {Rowe}, {Batalha}, {Borucki}, {Koch}, {Dunham}, {Gautier}, {Van Cleve},
  {Cochran}, {Latham}, {Lissauer}, {Torres}, {Brown}, {Gilliland}, {Buchhave},
  {Caldwell}, {Christensen-Dalsgaard}, {Ciardi}, {Fressin}, {Haas}, {Howell},
  {Kjeldsen}, {Seager}, {Rogers}, {Sasselov}, {Steffen}, {Basri},
  {Charbonneau}, {Christiansen}, {Clarke}, {Dupree}, {Fabrycky}, {Fischer},
  {Ford}, {Fortney}, {Tarter}, {Girouard}, {Holman}, {Johnson}, {Klaus},
  {Machalek}, {Moorhead}, {Morehead}, {Ragozzine}, {Tenenbaum}, {Twicken},
  {Quinn}, {Isaacson}, {Shporer}, {Lucas}, {Walkowicz}, {Welsh}, {Boss},
  {Devore}, {Gould}, {Smith}, {Morris}, {Prsa}, {Morton}, {Still}, {Thompson},
  {Mullally}, {Endl}, \& {MacQueen}}]{howard2012}
{Howard}, A.~W., {Marcy}, G.~W., {Bryson}, S.~T., {et~al.} 2012, \apjs, 201, 15

\bibitem[{{Huber} {et~al.}(2013){Huber}, {Chaplin}, {Christensen-Dalsgaard},
  {Gilliland}, {Kjeldsen}, {Buchhave}, {Fischer}, {Lissauer}, {Rowe},
  {Sanchis-Ojeda}, {Basu}, {Handberg}, {Hekker}, {Howard}, {Isaacson},
  {Karoff}, {Latham}, {Lund}, {Lundkvist}, {Marcy}, {Miglio}, {Silva Aguirre},
  {Stello}, {Arentoft}, {Barclay}, {Bedding}, {Burke}, {Christiansen},
  {Elsworth}, {Haas}, {Kawaler}, {Metcalfe}, {Mullally}, \&
  {Thompson}}]{huber2013}
{Huber}, D., {Chaplin}, W.~J., {Christensen-Dalsgaard}, J., {et~al.} 2013,
  \apj, 767, 127

\bibitem[{{Johnson} {et~al.}(2010){Johnson}, {Aller}, {Howard}, \&
  {Crepp}}]{johnson2010}
{Johnson}, J.~A., {Aller}, K.~M., {Howard}, A.~W., \& {Crepp}, J.~R. 2010,
  \pasp, 122, 905

\bibitem[{{Jones} {et~al.}(2016){Jones}, {Jenkins}, {Brahm}, {Wittenmyer},
  {Olivares E.}, {Melo}, {Rojo}, {Jord{\'a}n}, {Drass}, {Butler}, \&
  {Wang}}]{jones2016}
{Jones}, M.~I., {Jenkins}, J.~S., {Brahm}, R., {et~al.} 2016, \aap, 590, A38

\bibitem[{{Kunitomo} {et~al.}(2011){Kunitomo}, {Ikoma}, {Sato}, {Katsuta}, \&
  {Ida}}]{kunitomo2011}
{Kunitomo}, M., {Ikoma}, M., {Sato}, B., {Katsuta}, Y., \& {Ida}, S. 2011,
  \apj, 737, 66

\bibitem[{{Lillo-Box} {et~al.}(2014){Lillo-Box}, {Barrado}, {Henning},
  {Mancini}, {Ciceri}, {Figueira}, {Santos}, {Aceituno}, \&
  {S{\'a}nchez}}]{lillo-box2014}
{Lillo-Box}, J., {Barrado}, D., {Henning}, T., {et~al.} 2014, \aap, 568, L1

\bibitem[{{Liu} {et~al.}(2013){Liu}, {Guillochon}, {Lin}, \&
  {Ramirez-Ruiz}}]{liu2013}
{Liu}, S.-F., {Guillochon}, J., {Lin}, D.~N.~C., \& {Ramirez-Ruiz}, E. 2013,
  \apj, 762, 37

\bibitem[{{Lloyd}(2013)}]{lloyd2013}
{Lloyd}, J.~P. 2013, \apjl, 774, L2

\bibitem[{{Lopez} \& {Fortney}(2016)}]{lopez2016}
{Lopez}, E.~D., \& {Fortney}, J.~J. 2016, \apj, 818, 4

\bibitem[{{MacLeod} {et~al.}(2018){MacLeod}, {Cantiello}, \&
  {Soares-Furtado}}]{macleod2018}
{MacLeod}, M., {Cantiello}, M., \& {Soares-Furtado}, M. 2018, \apjl, 853, L1

\bibitem[{{Mayor} \& {Queloz}(1995)}]{mayor1995}
{Mayor}, M., \& {Queloz}, D. 1995, \nat, 378, 355

\bibitem[{{Ogilvie} \& {Lin}(2004)}]{ogilvie2004}
{Ogilvie}, G.~I., \& {Lin}, D.~N.~C. 2004, \apj, 610, 477

\bibitem[{{Reffert} {et~al.}(2015){Reffert}, {Bergmann}, {Quirrenbach},
  {Trifonov}, \& {K{\"u}nstler}}]{reffert2015}
{Reffert}, S., {Bergmann}, C., {Quirrenbach}, A., {Trifonov}, T., \&
  {K{\"u}nstler}, A. 2015, \aap, 574, A116

\bibitem[{{Sato} {et~al.}(2005){Sato}, {Kambe}, {Takeda}, {Izumiura}, {Masuda},
  \& {Ando}}]{sato2005}
{Sato}, B., {Kambe}, E., {Takeda}, Y., {et~al.} 2005, \pasj, 57, 97

\bibitem[{{Schlaufman} \& {Winn}(2013)}]{schlaufman2013}
{Schlaufman}, K.~C., \& {Winn}, J.~N. 2013, \apj, 772, 143

\bibitem[{{Van Eylen} {et~al.}(2016){Van Eylen}, {Albrecht}, {Gandolfi}, {Dai},
  {Winn}, {Hirano}, {Narita}, {Bruntt}, {Prieto-Arranz}, {Bejar}, {Nowak},
  {Lund}, {Palle}, {Ribas}, {Sanchis-Ojeda}, {Yu}, {Arriagada}, {Butler},
  {Crane}, {Handberg}, {Deeg}, {Jessen-Hansen}, {Johnson}, {Nespral}, {Rogers},
  {Ryu}, {Shectman}, {Shrotriya}, {Slumstrup}, {Takeda}, {Teske}, {Thompson},
  {Vanderburg}, \& {Wittenmyer}}]{vaneylen2016}
{Van Eylen}, V., {Albrecht}, S., {Gandolfi}, D., {et~al.} 2016, ArXiv e-prints,
  arXiv:1605.09180

\bibitem[{{Vanderburg} {et~al.}(2015){Vanderburg}, {Latham}, {Buchhave},
  {Bieryla}, {Berlind}, {Calkins}, {Esquerdo}, {Welsh}, \&
  {Johnson}}]{vanderburg2015}
{Vanderburg}, A., {Latham}, D.~W., {Buchhave}, L.~A., {et~al.} 2015, ArXiv
  e-prints, arXiv:1511.07820

\bibitem[{{Villaver} \& {Livio}(2009)}]{villaver2009}
{Villaver}, E., \& {Livio}, M. 2009, \apjl, 705, L81

\bibitem[{{Villaver} {et~al.}(2014){Villaver}, {Livio}, {Mustill}, \&
  {Siess}}]{villaver2014}
{Villaver}, E., {Livio}, M., {Mustill}, A.~J., \& {Siess}, L. 2014, \apj, 794,
  3

\end{thebibliography}


\begin{thebibliography}{}

\bibitem[Akeson et al.(2013)]{akeson2013} Akeson, R.~L., Chen, X., Ciardi, D. et al.\ 2013, \pasp, 125, 989

\bibitem[Barclay et al.(2015)]{barclay2015} Barclay, T., Endl, M., Huber, D. et al.\ 2015, \apj, 800, 46

\bibitem[Barclay et al.(2018)]{barclay2018} Barclay, T., Pepper, J., \& Quintana, E.~V.\ 2018, ArXiv e-prints, arXiv:1804:05050

\bibitem[Bodenheimer et al.(2001)]{bodenheimer2001} Bodenheimer, P., Lin, D.~N.~C., \& Mardling, R.~A.\ 2001, \apj, 548, 466

\bibitem[Butler et al.(1996)]{butler1996} Butler, R.~P., Marcy, G.~W., Williams, E. et al.\ 1996, \pasp, 108, 500

\bibitem[Campante et al.(2016)]{campante2016} Campante, T.~L., Schofield, M., Kuszlewicz, J.~S. et al.\ 2016, \apj, 830, 138

\bibitem[Chen \& Kipping(2017)]{chen2017} Chen, J,. \& Kipping, D.\ 2017, \apj, 834, 17

\bibitem[Dawson \& Johnson(2018)]{dawson2018} Dawson, R.~I. \& Johnson, J.~A., ArXiv e-prints, arXiv:1801.06117

\bibitem[Dobbs-Dixon et al.(2004)]{dobbs-dixon2004} Dobbs-Dixon, I., Lin, D.~N.~C., \& Mardling, R.~A.\ 2004, \apj, 610, 464

\bibitem[Dressing \& Charbonneau(2015)]{dressing2015} Dressing, C.~D. \& Charbonneau, D.\ 2015, \apj, 807, 45

\bibitem[Eastman et al.(2013)]{eastman2013} Eastman, J., Gaudi, B.~S., \& Agol, E.\ 2013, \pasp, 125, 83

\bibitem[Fuller(2017)]{fuller2017} Fuller, J.\ 2017, \mnras, 472, 1538

\bibitem[Fulton et al.(2018)]{fulton2018} Fulton, B.~J., Petigura, E.~A., Blunt, S., \& Sinukoff, E.\ 2018, ArXiv e-prints, arXiv:1801.01947

\bibitem[Fulton et al.(2017)]{fulton2017} Fulton, B.~J., Petigura, E.~A., Howard, A.~W., et al.\ 2017, ArXiv e-prints, arXiv:1703.10375

\bibitem[Gallet et al.(2017)]{Gallet2017} Gallet, F., Bolmont, E., Mathis, S., Charbonnel, C., \& Amard, L.\ 2017, ArXiv e-prints, arXiv:1705.10164

\bibitem[Goldreich \& Soter(1966)]{goldreich1966} Goldreich, P., \& Soter, S.\ 1966, Icarus, 5, 375

\bibitem[Grunblatt et al.(2016)]{grunblatt2016} Grunblatt, S.~K., Huber, D., Gaidos, E., et al.\ 2016, \aj, 152, 185

\bibitem[Grunblatt et al.(2017)]{grunblatt2017} Grunblatt, S.~K., Huber, D., Gaidos, E., et al.\ 2017, \aj, 154, 254

\bibitem[Howard et al.(2012)]{howard2012} Howard, A.~W., Marcy, G.~W., Bryson, S.~T., et al.\ 2012, \apjs, 201, 15

\bibitem[Huber et al.(2013)]{huber2013} Huber, D., Chaplin, W.~J., Christensen-Dalsgaard, J., et al.\ 2013, \apj, 767, 127

\bibitem[Huber et al.(2016)]{huber2016} Huber, D., Bryson, S.~T., Haas, M.~R. et al.\ 2016, \apjs, 224, 2

\bibitem[Hut(1981)]{hut1981} Hut, P.\ 1981, A\&A, 99, 126

\bibitem[Johnson et al.(2010)]{johnson2010} Johnson, J.~A., Aller, K.~M., Howard, A.~W., \& Crepp, J.~R.\ 2010, \pasp, 122, 905

\bibitem[Jones et al.(2016)]{jones2016} Jones, M.~I., Jenkins, J.~S., Brahm, R. et al.\ 2016, A\&A, 590, A38

\bibitem[Kunitomo et al.(2011)]{kunitomo2011} Kunitomo, M., Ikoma, M., Sato, B., Katsuta, Y., \& Ida, S.\ 2011, \apj, 737, 66

\bibitem[Lillo-Box et al.(2016)]{lillo-box2016} Lillo-Box, J., Barrado, D., Correia, A.~C.~M.\ 2016, A\&A, 589, A124

\bibitem[Lillo-Box et al.(2014)]{lillo-box2014} Lillo-Box, J., Barrado, D., Moya, A., et al.\ 2014, A\&A, 562, A1090.49

\bibitem[Lopez \& Fortney(2016)]{lopez2016} Lopez, E.~D., \& Fortney, J.~J.\ 2016, \apj, 818, 4

\bibitem[MacLeod et al.(2018)]{macleod2018} MacLeod, M., Cantiello, M., \& Soares-Furtado, M.\ 2018, \apj, 853, L1

\bibitem[Miller et al.(2009)]{miller2009} Miller, N., Fortney, J.~J., \& Jackson, B.\ 2009, \apj, 702, 1413

\bibitem[Morton et al.(2016)]{morton2016} Morton, T.~D., Bryson, S.~T., Coughlin, J.~L., et al.\ 2016, \apj, 822, 86

\bibitem[Niedzielski et al.(2016)]{niedzielski2016} Niedzielski, A., Villaver, E., Nowak, G., et al.\ 2016, A\&A, 589, L1


\bibitem[Patra et al.(2017)]{patra2017} Patra, K.~C., Winn, J.~N., Holman, M.~J., et al.\ 2017, ArXiv e-prints, arXiv:1703.06582

\bibitem[Petigura et al.(2013)]{petigura2013} Petigura, E.~A., Howard, A.~W., \& Marcy, G.~W.\ 2013, Proceedings of the National Academy of Science, 110, 19273

\bibitem[Petigura et al.(2017)]{petigura2017b} Petigura, E.~A., Sinukoff, E., Lopez, E.~D., et al.\ 2017a, \aj, 153, 142


\bibitem[Pollacco et al.(2006)]{pollacco2006} Pollacco, D.~L., Skillen, I., Collier Cameron, A., et al.\ 2006, \pasp, 118, 1407

\bibitem[Santerne et al.(2016)]{santerne2016} Santerne, A., Moutou, C., Tsantaki, M., et al.\ 2016, A\&A, 587, A64

\bibitem[Sato et al.(2005)]{sato2005} Sato, B., Kambe, E., Takeda, Y., et al.\ 2005, \pasj, 57, 97

\bibitem[Simpson(1951)]{simpson1951} Simpson, P.~B.\ 1951, The Annals of Mathematical Statistics, 22, 476

\bibitem[Stephens(1974)]{stephens1974} Stephens, M.~A.\ 1974, Journal of the American Statistical Association, 69, 730

\bibitem[Storch \& Lai(2014)]{storch2014} Storch, N.~I., \& Lai, D.\ 2014, \mnras, 438, 1526

\bibitem[Sullivan et al.(2015)]{sullivan2015} Sullivan, P.~W., Winn, J.~N., Berta-Thompson, Z.~K. et al.\ 2015, \apj, 809, 77

\bibitem[Van Eylen \& Albrecht(2015)]{vaneylen2015} Van Eylen, V., \& Albrecht, S.\ 2015, \apj, 808, 126

\bibitem[Van Eylen et al.(2016)]{vaneylen2016} Van Eylen, V., Albrecht, S., Gandolfi, D., et al.\ 2016, ArXiv e-prints, arXiv:1605.09180

\bibitem[van Sluijs \& Van Eylen(2018)]{vansluijs2018} van Sluijs, L., \& Van Eylen, V.\ 2018, \mnras, 474, 4603

\bibitem[Veras(2016)]{veras2016} Veras, D.\ 2016, Royal Society Open Science, 3, 150571

\bibitem[Villaver \& Livio(2009)]{villaver2009} Villaver, E., \& Livio, M.\ 2009, \apj, 705, L81

\bibitem[Villaver et al.(2014)]{villaver2014} Villaver, E., Livio, M., Mustill, A.~J., \& Siess, L.\ 2014, \apj, 794, 3

\bibitem[Xie et al.(2016)]{xie2016} Xie, J.-W., Dong, S., Zhu, Z., et al.\ 2016, Proceedings of the National Academy of Science, 113, 11431

\bibitem[Zahn(1977)]{zahn1977} Zahn, J.-P.\ 1977, A\&A, 57, 383

\end{thebibliography}

\end{document}